# Building trust in digital policing: A scoping review of community policing apps.

**Corresponding author: Camilla Elphick camilla.elphick@open.ac.uk[1]**
Richard Philpot r.philpot@lancaster.ac.uk[2]; Min Zhang min.zhang@open.ac.uk[1]; Avelie Stuart a.stuart@exeter.ac.uk[3]; Zoe Walkington z.walkington@open.ac.uk[1]; Lara Frumkin lara.frumkin@open.ac.uk[1]; Graham Pike graham.pike@open.ac.uk[1]; Kelly Gardner kelly.gardner@thamesvalley.pnn.police.uk[3] ; Mark Lacey mark.lacey@thamesvalley.pnn.police.uk[3]; Mark Levine mark.levine@lancaster.ac.uk [2]; Blaine Price b.a.price@open.ac.uk[1]; Arosha Bandara arosha.bandara@open.ac.uk[1]; Bashar Nuseibeh bashar.nuseibeh@open.ac.uk[1&5]

[1]The Open University, Walton Hall, Kents Hill, Milton Keynes, MK7 6AA
[2]Lancaster University, Bailrigg, Lancaster, LA1 4YW
[3]University of Exeter, Stocker Road, Exeter, EX4 4PY
[4]Thames Valley Police, Witan Gate, Milton Keynes, MK9 2DS
[5]Lero - The Irish Software Research Centre, University of Limerick, Limerick, Ireland

Perceptions of police trustworthiness are linked to citizens' willingness to cooperate with police. Trust can be fostered by introducing accountability mechanisms, or by increasing a shared police/citizen identity, both which can be achieved digitally. Digital mechanisms can also be designed to safeguard, engage, reassure, inform, and empower diverse communities. We systematically scoped 240 existing online citizen-police and relevant third-party communication apps, to examine whether they sought to meet community needs and policing visions. We found that 82% required registration or login details, 55% of those with a reporting mechanism allowed for anonymous reporting, and 10% provided an understandable privacy policy. Police apps were more likely to seek to reassure, safeguard and inform users, while third-party apps were more likely to seek to empower users. As poorly designed apps risk amplifying mistrust and undermining policing efforts, we suggest 12 design considerations to help ensure the development of high quality/fit for purpose Police/Citizen apps.

**Funding:** This work was supported by the Citizen Forensics project, funded by the UK EPSRC (EP/R033862/1 and EP/R013144/1), and Science Foundation Ireland (SFI 13/RC/2094, 16/SP/3804).

**Declarations of interest:** none

Police/Citizen apps.Policing 101 (U.S. Department of Justice, n.d.) states that:

> **Positive police-community relationships are essential to maintaining public safety and order. These relationships help to reduce fear and biases, and build mutual understanding and trust between the police and community (page 9).**

The importance of these relationships are based upon Peel's (1929) nine principles of policing (cited in The Home Office n.d., The Law Enforcement Action Partnership n.d.), the seventh of which conceives of the police also as civilians and civilians also as police. However, there is some debate as to the extent to which US police are connected to these principles (Adegbile 2017), as there are more than 18,000 police departments in the US that are subject to different laws and codes (U.S. Department of Justice n.d.). In the UK, Peelian principles have also influenced policing (The Home Ofice n.d), but the sections of the Policing Vision 2025 of relevance to community policing are focused on 'protecting and reassuring communities' (NPCC 2015). Whether community policing strategies seek to reduce fear and bias, or protect and reassure communities, policing efforts can be hampered by a range of factors including: power imbalances and inequality (e.g. Gasper 2012); prejudice (Kochel et al. 2011); and by a lack of clear or transparent policies or procedures (see Brucato 2015, for a commentary). These issues contribute to mistrust (Ray et al 2017) and an asymmetry in citizen-police collaboration (e.g. Fulla and Welch 2002) [1].

Goldsmith (2005) described trust as the ways that interactions and experiences contribute to expectations about future treatment. Based on three presumptions, summarised by Six (2003) as benevolence, dedication and ethics, Goldsmith lists nine behaviours that contribute to mistrust in police. These are: neglect; indifference; incompetence; discrimination; brutality; venality; extortion; intimidation; and excessive force. Goldsmith also proposed that trust can be increased with accountability, and listed three key actions of information, influence, and control (identified by Six 2003) to achieve this.

---

[1] This research is of value internationally. However, the authors were based in the UK, and most of the apps in the final sample were hosted in the US, so we focused on the policing principles and missions of the US and the UK.

It is in the interest of police and citizens to foster trust, as there is a robust association between police trustworthiness (Goldsmith 2005), fairness and/or perceived legitimacy (Jackson, et al. 2012), and cooperation with police (see Cao 2014, for a commentary) Fostering trust has become urgent in the advent of the widespread use of social media, as civilians are able to share police data (see Bradford & Jackson, 2010, for a review of trust in police; Walsh & O'Connor, 2018, for a review of policing and social media, and Sambrook, 2012, for a commentary on trust in the digital age). For instance, genuine mistakes, such as the shooting of Jean Charles de Menezes (2005); incidents of inappropriate behaviour (Weitzer 2006); incompetency or coverups (e.g. the death of Christian Andreacchio; Tenderfoot TV & Black Mountain Media 2019); or police brutality (e.g. the killings of Freddie Gray, Undisclosed 2017; and George Floyd, BBC 2020) are accessible online. Therefore, tensions between police and some communities are high (Upton Patton et al. 2016).

One way to foster trust is to introduce accountability mechanisms for citizens to collect evidence, file complaints, raise community awareness, verify police performance (and corrective action), and to support citizens while doing so (Goldsmith 2005). Another approach might be to build community identities (Reicher & Hopkins 2000) and shared identities between police and citizens, as this can increase social solidarity between them (Jackson & Bradford 2010). Citizens are also more willing to cooperate in investigations and disclose information when they identify with the police and feel included in what the group represents (Bradford 2014). Shared identification with the police should therefore promote citizen information disclosure and cooperation, which may assist police investigations.

Digital technologies help to create shared identities (Hartz-Karp et al. 2010) by creating online communities (e.g. Mumsnet n.d.) or allowing individuals to share common goals (e.g. Movember n.d.). They can involve interactive "smart" systems that are scalable (Chaudhari et al. 2018) to diverse needs and interests of individuals or communities (Mills 2011). They can also involve people in 'citizen forensics' such as sharing evidence, advocacy, or 'web sleuthing'. This is a practice where people use publicly-available resources to conduct amateur investigations, in an attempt to solve crimes (e.g. Yardley et al. 2018), such as 'Facebook identifications' (Mack and Sampson 2013) before attending an official police lineup, or searching for missing people (Cruz-Santiago 2017). There are also an increasing number of crime-solving websites (e.g. crimemuseum.org), forums (e.g.WebSleuths.com), and podcasts (e.g. Undisclosed-podcast.com), which can appeal to people's interest in solving crime. Web sleuthing justifiably raises concerns about vigilantism (e.g. Campbell 2016). However, a recent literature review by Yardley et al. found that under four percent of web sleuthing activities were related to concerns about vigilantism, while almost ninety one percent were beneficial (e.g. content co-creation and exchange). This suggests that web sleuthing can be helpfully integrated into criminal justice systems (Huey et al. 2013), for instance via Police Support Officers (Home Office 2016), provided that digital technologies are designed to minimise vigilantism. Steps also need to minimise the inclusion of erroneous web sleuthing information, which can interfere with investigations, erode trust, and make police cautious about providing web sleuthing mechanisms (Huey et al. 2013).

Digital technologies also allow "digital citizens" to participate in and challenge the practices of authorities and institutions (Mossberger et al. 2007). There is huge potential to provide forensically assured digital accountability mechanisms, as these can be programmed to be inclusive, transparent, fair, and available at any time. However, they need to be designed in a way that amplifies the overarching police mission, whilst having an architecture that is flexible enough to align with specific demographics. They also need to be managed by police officers who are technically skilled and motivated to interact with them. Thus, digital accountability mechanisms have the potential to transform traditional models of public service, governance and civic engagement, potentially motivating individuals in different communities to engage in things that are relevant to them.

In a report for the US Office of Community Oriented Policing Services, Cordner and Perkins (2013) demonstrated that digital technologies are already being used for citizen-police communication in an attempt to enhance and support existing community policing However, as these technologies have not been built on systematic research, work is needed to understand how to implement them effectively (for instance, Hu and Lovrich (2020) have since explored the relationship between social media and policing). Currently, little is understood about how they are used; whether they seek to reassure citizens (that police are transparent, accountable, and open to feedback); whether they seek to empower citizens (so that they can hold authorities to account); whether they seek to engage citizens (in an attempt to create a shared identity); or whether they just disseminate top-down information. To maximise effective digital collaboration between citizens and police, greater understanding of how digital evidence and intelligence is gathered, used, and appropriated, is required.

The aims of this study are to systematically scope existing online citizen-police communication apps, and to examine whether they seek to: enhance crime prevention and response; meet community needs; engage, include and empower "digital" citizens; foster shared identities; and whether they are forensically assured, private, and contextually appropriate. The study focuses on similarities and differences between apps hosted by police and those hosted by third parties (including charities, companies, and community groups). Specifically, it asks whether the apps:

> Protect user data
> Seek to reassure users
> Seek to safeguard users
> Seek to empower or engage users
> Seek to share information with users
> Are navigable and presentable

The study concludes by providing design considerations for the development of future Police-Citizen apps, based upon the present research.

# Methods

## Research Design

Informed by the Preferred Reporting Items for Systematic Reviews and Meta-Analyses (PRISMA) guidelines (Moher et al. 2010)[2], we used keywords to search for police/citizen apps that contained an integrated two-way communication mechanism. On the final sample, we scored fifteen measures, focusing on Data Protection and App Features.

--------------------------

[2] PRISMA helps to ensure transparent and complete reporting when conducting systematic reviews, with a checklist of items to include (Liberati, Altman, Tetzlaff, Mulrow, Gøtzsche, Ioannidis et al., 2009).

**Apps Sources**

To systematically retrieve relevant apps, a computer search of the Google Play store (n.d.) was conducted[3]. Initial computer searches were conducted by one author with computing expertise according to keywords agreed upon by all authors. Subsequent inclusion/exclusion steps were led by two other authors.

**Computer Search**

The Google Play store was searched using combinations of the following keywords:

Community/community-driven/neighbo(u)rhood/smart+policing/app/engagement/collaboration/communication. Community/neighbo(u)rhood+crime+evidence/witness/prevention/reporting/justice /+app

A standalone scraping tool (Parsehub Standard Version n.d.) was used to extract the apps. The search was conducted between 1st July to 4th July (inclusive), 2019, by authors from the computing department of university. Following the steps taken by them, a final list of 1207 apps was identified. At this point, authors from the psychology department followed inclusion/exclusion steps before running the analyses. The search was conducted within a limited time, to control for the ephemeral nature of apps. Indeed, a few apps became inaccessible in the time between collection and analysis due to broken URLs. Thus, all analyses in the present study relate only to the way existing apps functioned at the time of analysis (16th September – 15th October, 2019, inclusive).

**Exclusion and Inclusion Criteria**

The aim was to evaluate apps that encouraged interactive dyadic (two-way) communication between police and citizens (or between citizens and citizens about police or community concerns). As it was not possible to detect whether apps included mechanisms that allowed for dyadic conversations just by reading the information provided on Google Play store listing webpages, the presence of some kind of integrated communication mechanism was designated the minimum requirement for inclusion. Apps that did not meet this threshold were excluded. When it came to the stakeholders concerned (e.g. police or digital citizens), we used information provided on Google Play store listing webpages. We only included apps either hosted by police or related to citizen concerns. Thus, inclusion/exclusion criteria were shaped by the presence or absence of an interactive communication mechanism (e.g. reporting, feedback, or messaging), and the stakeholders concerned (for full inclusion/exclusion steps, see Appendix A, for app locations, see Appendix B). Our study focuses on mobile applications rather than web sites to ensure like-for-like comparisons of the functionalities provided. Including web sites would have made it harder to identify a coherent set of design considerations, as police websites serve a wider range of purposes.

_______________________

[3] We focused on the Android operating system because 76% of global mobile device market share uses Android (Statcounter, n.d.) Also, with >1M more apps than Apple iOS, the Android app store is likely to have a greater selection of relevant community policing apps. Indeed, curated listings of policing apps do not identify any apps that are solely developed for iOS devices (Rasmussen, n.d.).

The final sample included 240 apps. These were installed on Android One Mi A2 Lite mobile phones. Research phones were registered to research google accounts, to protect the privacy of the authors when using the apps.

**Measurements**

For the final 240 apps we scored 15 measures, some of which were taken from a previous website scoping review (Moore and Harrison, 2018), some which were provided by Google Play, and some which were driven by the examination of the apps themselves. For instance, some apps included statistics, mission statements, safeguarding tips, or events, but others did not, so these items were divided into six different measures, under the heading 'app functions'. The definition and source of each measure is described below.

*Measures identified by Moore & Harrison (2018):*

**Host:** This measure was inspired by Moore and Harrison's measure 'author' (2018). We changed the name to reflect the fact that apps are commonly designed by developers who do not host the final product[4]. Hosts were coded into 'third-party' (private businesses, or community groups e.g. neighbourhood watch, charities, NGOs); & 'police' (police, sheriffs etc.).

**Navigation:** This measure determined how easy it was for users to find their way around the app, and/or how well maintained it was. Apps with a clear menu or index that linked to each page were assigned 2 points. Apps that needed several clicks (or were unclear/confusing) to access some pages scored 1. Apps scored 0 if they were difficult to navigate, e.g. lack of menu, confusing (or hidden) links, navigation loops, or if they needed a search option. The range was from 0-2 points.

**Presentation:** Clear, uncluttered apps (e.g. a balance between text and pictures) were assigned 2 points. A score of 1 was given to apps with mediocre presentation (e.g. images that made poor use of space). Confusing and cluttered apps scored 0 (e.g. too much information on a page). The range was from 0-2 points.

*Measures identified in the apps:*

**Data Protection**

**Informed consent:** This measure was inspired by Moore and Harrison's measure 'readability' (2018). It was included to determine whether users could access a privacy policy before installing the app (on the app landing page), and whether it was understandable. Failure to make policies readable to the average user is discriminatory, and of ethical concern, as users will not have provided informed consent if they do not understand the policy. We tested whether privacy policies were readable by calculating a Flesch reading ease score with an online calculator (Felsch n.d.). Scores range from 0-100, and scores of 60-70 are considered acceptable.

**Inclusivity:** It is important for police apps to be as inclusive as possible, so that marginalised groups can access them more readily. Apps that did not require data permissions, login or registration, scored 3 points (fully inclusive); those that required some personal data (e.g. location) to access certain functions (partially inclusive) scored 2 points; those that required users to create an account or register to access them (partially exclusive) scored 1 point; those that required additional permissions or membership to install them, or had fees that were not mentioned on the app website, (fully exclusive) scored 0 points. There was a range from 0-3 points.

---

[4] App designs are not only driven by policing concerns. External developers have their own interests in the app and the data generated.

**Anonymity:** Anonymous reporting is the standard approach taken in the sector, including by Crimestoppers (n.d). Anonymity is important when reporting sensitive issues (e.g. crime) and safeguarding citizens from possible retaliation, so it is important that apps which include reporting mechanisms also protect user's anonymity when reporting. We explored this by seeing whether apps with reporting mechanisms also included *anonymous reporting options*. For an app with a reporting mechanism to be categorised as 'anonymous', anonymity options had to be (i) made explicit before a user made a report (as it is important for users to know what will happen to their data prior to reporting), and (ii) provide anonymous options. Apps that met both these criteria scored 1 point, otherwise they scored 0 points[5].

**App Features**

**Reassuring:** We focused on Six's key actions (2003) and Goldsmith's behaviours (2005). These were information (e.g. a mission statement or feedback mechanism); influence (e.g. response to citizen feedback); discrimination (acknowledgement of or commitment to reducing discrimination); and brutality (e.g. citizen rights). This measure also reflected the idea of accountability mechanisms (Goldsmith 2005) and was key to the reassurance element of the Policing Vision 2025 (NCPP 2015).

If apps included either procedural policies, freedom of information, or a mission statement, they scored 1 point, if they did not, they scored 0 points; if apps claimed to follow up user input (e.g. communicated with a citizen who had submitted a crime tip) they scored 1 point, if not they scored 0 points; if they claimed to provide users with a feedback mechanism (e.g.

---

[5] We did not test actual anonymity, as this would have involved submitting a fake police report. Thus, it was impossible to know whether statements about anonymity were carried out in practice. Thus, this measure only gave an impression of the extent to which the app hosts were taking anonymity into account.

complaint) they scored 1 point, if not they scored 0 points; if they informed users of their rights or demonstrated a commitment to reducing prejudice or bias they scored 1 point, if not they scored 0 points. There was a range of 0-4 points.

**Safeguarding:** This measure reflected the safety element of the Policing Vision 2025 (NCPP 2015). If apps provided alerts or warnings (e.g. crimes or natural disasters), wanted 'posters', safety tips (e.g. burglar prevention), tools (e.g. compasses, maps), crime maps, house watch schemes (e.g. police 'drives-by'), property logging, checking in schemes (e.g. on vulnerable adults) they scored a point for each. Also, if apps provide an integrated means to get immediate response to an emergency (e.g. a 999 button) they scored 1 point, if not they scored 0 points. There was a range from 0-9 points.

**Empowering:** This measure reflected the idea of using accountability mechanisms to raise community awareness (Goldsmith 2005). For this, we focused on Six's key action (2003) of control (sharing responsibility, delegating or collaborating with citizens). If apps provided means for communities to respond to SOS (e.g. a distressed user alerting other users); to notify others about risk (e.g. a crime or natural disaster); or raised awareness, rights, resources (e.g. charity funding pages), they scored 1 point for each. The range was 0-3 points.

**Engaging:** This measure evaluated the extent to which apps sought to create a shared citizen/police identity (Jackson & Bradford, 2010) with shared values, including the notion of police being citizens as well as law enforcers (The Home Office n.d.). If apps provided social media (e.g. Twitter), chat options (e.g. a wall or forum), recruitment, events, a gallery, membership or collaboration, training (e.g. cadets), community interaction (e.g. 'rides-along'), they scored a point. The range was from 0-8 points.

**Information Sharing:** This measure described the type of information shared, and whether it was accurate and timely (Six 2003). If apps provided information: statistics; terms and definitions; reports; news; FAQs; they scored 1 point for each. Also, if the app provided document services (e.g. applications or online forms) they scored 1 point, if not they scored 0. The range was from 0-6 points.

**Surveillance and Tracking:** This measure was included as providing information about sex offenders can lead to vigilantism (Cubellis et al. 2018). Some apps provided users with ways to track or surveil individuals (e.g. sex offender maps). These were deemed ethically unsound, and with the potential to incite vigilantism. Therefore, if these were present, the app scored 0 points, if they were not present, the app scored 1 point.

The range of possible scores was therefore 0-40 points in total.

To evaluate the systematic reliability of the measures, two raters independently scored a subsample of 35 apps (20.47% of the total sample). We calculated Gwet's $AC_1$ and $AC_2$ coefficients to assess interrater agreement (Gwet 2014). All agreements were between .79 and 1 indicating excellent reliability (Fleiss 1981) (Appendix C).

*Measures provided by Google Play:*

**App_ratings:** These were average user ratings (with a possible range between 1 and 5). The minimum score users could select was 1 (for apps users considered to be poor), and the maximum score they could select was 5 (for apps they considered to be excellent)[6]. This was the sole user experience measure.

## Results

### Data Protection

The present research was interested in differences between police and third-party apps, so we analysed the data protection measures as a function of the app host (police; third-party). The null hypotheses were that there would be no difference in i) informed consent; ii) inclusivity; or iii) anonymity, between police and third-party apps. When it came to informed consent, we included data from the 171 apps (as we could access the data of exclusive apps without installing them), but for the remaining measures we only analysed data from inclusive apps, as only these would be accessible to citizens without logging in (n = 64).

───────────────────────

[6] Google changed their rating mechanism in August, 2019. The present data was analysed after this change, but was collected before the change, so it might not reflect current Google ratings.

**Informed Consent**

We tested whether users could access a privacy policy before installing the app, and whether that policy was understandable to the average user, as users will not have provided informed consent if they do not understand the policy. The privacy policies (that explained how user data would be collected and used) were usually found on the landing page of each app, as it was at this point that users could decide whether to install the app or not. We were interested in determining how readable these privacy policies were. We used Flesch (n.d.) to test the readability of each privacy policy, as it provides clear labels about how easy a document is to read, as well as providing scores.

When calculating reading ease, the mean readability score was 39.18 ($SD$ = 8.87). This indicates that the policies were largely 'difficult to read' for someone without a university level education. A reading ease score of 60-70 is considered acceptable or 'average', but only three apps (1.75%) with privacy policies were within this acceptable range: We Are All Police (67.5); Rutland Neighbourhood Watch (65.9); and Your999 (64.7). 35.67% either provided no privacy policy or there was no access to it.

To determine the relationship between *host* and *informed consent*, a chi-square was conducted, but no significant relationship was found, $\chi^2$ = 5,70, $p$ = .06: it made no difference who hosted the apps (third-party; police) when it came to providing a readable privacy policy. However, as the results were marginally non-significant, we inspected them. They revealed that apps hosted by police had *slightly* less readable privacy policies ($M$ = 38.71, $SD$ = 6.92) than those hosted by third-parties ($M$ = 39.62, $SD$ = 10.44). In short, while it is possible to make privacy policies readable by the average user, the overwhelming majority of hosts fail to do so, compromising the privacy of users who may not understand to what they are agreeing before using the app.

**Inclusivity**

When trying to install the 240 apps for scoring, 69 could not be installed (referred to as 'fully exclusive' apps). The reasons for this included a broken URL, ineligibility (e.g. accessible to first responders only), geographical location (detected from IP addresses), or because the app required a fee or special software to install it. However, 171 apps were installed successfully. Of these, 104 required the user to create an account or to register to install them ('partially exclusive'). These were not included in the analyses below, as we only analysed apps that we could access without logging in. This resulted in 67 inclusive apps, of which 24 required personal data to access certain functions ('partially inclusive'). By the time of analysis, three of these 24 were no longer available (broken URLs), so there remained only 21 partially inclusive apps. The remaining 43 were 'fully inclusive'. This resulted in a final sample of 64 apps.

To determine the relationship between *host* and *inclusivity,* a chi-square was conducted, and a significant relationship was found, $\chi^2 = 28.37$, $p < .001$: the proportion of police apps that were fully inclusive (40.96%) was significantly higher than the proportion of third-party apps (10.23%) that were fully inclusive (rather than partially inclusive). Thus, while considerably more police apps were fully inclusive than third-party apps, most were partially inclusive, making them inaccessible to some users.

### Anonymity

Anonymity is an important consideration when designing apps, which commonly collect data including IP addresses, geographical locations, or device information. Given the potentially sensitive nature of matters reported (e.g. crimes), or potential risk to users of reporting (e.g. retaliation), we were interested in data shared via communication mechanisms. Therefore, explored whether apps that included an integrated reporting mechanism also claimed to provide anonymous ways of reporting (user anonymity is discussed more widely in the discussion). We could only access the reporting mechanisms of the 64 inclusive apps. Of these 86% included an integrated reporting mechanism, of which 55% explicitly stated that they allowed for anonymous reporting *prior to* making a report.

When testing the relationship between *host* and *anonymity*, a fisher's exact test was not statistically significant, $\chi^2 = 0.02$, $p = .86$: police apps were no more (or less) likely to include anonymous reporting options (54.17%) than third-party ones (56.25%). Thus, while most apps provided some kind of reporting mechanism, only about half made it clear to users before reporting what kind of control they had over their anonymity. This is of particular concern in apps that also failed to provide readable privacy policies, leaving users unclear as to the risks they are taking with their data when reporting crimes.

## App Features

Next, we analysed relationships between *host* and *app features* (n = 64). The null hypotheses were that apps hosted by police or third-parties were no more or less likely to i) reassure, ii) safeguard, iii) empower, iv) engage, v) share information, or vi) surveil or track.

### Reassuring

There was a range of 0-4 points ($M = 0.91$; $SD = 0.86$). The three highest scoring apps were Horry County Police Department, RCIPS, and Largo Police (3 points each). These included a friendly welcome page with a link to their mission statement (Horry County); detailed information about providing feedback (RCIPS); and detailed information about their approach to bias (Largo Police). As shown in table 1, no significant relationship was found between *host* and *reassuring*, $F(1,63) = 1.40$, $p = .24$: police apps were no more or less likely to reassure than third-party apps. Thus, although building trust is important to policing, police apps were not sufficiently seeking to reassure users. On average, they only included about one of the four possible reassurance items.

### Safeguarding

There was a range of 0-9 points ($M = 2.15$; $SD = 1.62$). The two highest scoring apps were Frederick County Sheriff, and Eloy Police Department (6 points each). They included items such as vacation watch (checking vacant homes), 'wanted' suspects, emergency response (911), and safety tips (e.g. burglary prevention). A significant relationship was found between *host* and *safeguarding*, $F(1,63) = 4.36$, $p = .04$, $\eta^2 = .07$: apps hosted by police were more likely to provide safeguarding items than those hosted by third-parties. However, on average, they only included 2.39 items out of a possible nine.

### Empowering

There was a range of 0-3 points ($M = 0.45$; $SD = 0.62$). The four highest scoring apps were GATOR SAFE, Rutland Neighbourhood Watch, Hinckley Neighbourhood Watch, and LAPD Devonshire (2 points each). They included items such as 'Friend Walk', involving sending location data to a friend who can trigger an emergency call (GATOR SAFE); a comments 'wall', for alerting others to real-time incidents (Rutland and Hinckley Neighbourhood Watch); and neighbourhood watch schemes (LAPD Devonshire). A significant relationship was found between *host* and *empowering*, $F(1,62) = 5.30$, $p = .03$, $\eta^2 = .08$: third-party apps were more empowering than police apps, although on average, third-party apps only included 0.75 items out of a possible three.

### Engaging

There was a range of 0-8 points ($M = 2.16$; $SD = 1.53$). The five highest scoring apps were Yakima Police Department, DUBAI POLICE, Sebastian Police Department, El Cerrito Police Department, and Fellsmere Police Department (5 points each). They included items such as volunteering (Yakima Police Department), police museum tours (DUBAI POLICE), a gallery (Sebastian Police Department), rides-along (El Cerrito Police Department), and events (Fellsmere Police Department), as well as social media links. A significant relationship was found between *host* and *engaging*, $F(1,62) = 4.15$, $p < .05$, $\eta^2 = .06$: police apps were more engaging than third-party apps. However, on average, they only included 2.38 items out of a possible eight.

### Information Sharing

There was a range of 0-6 points ($M = 1.94$; $SD = 1.46$). The three highest scoring apps were El Cerrito Police Department, Los Angeles County Sheriff and Medicine Hat Police Service (5 points each). They included items such as FAQs (El Cerrito Police Department), statistics (Los Angeles County Sheriff), application forms (Medicine Hat Police Service). A significant relationship was found between *host* and *information sharing*, $F(1,62) = 6.11$, $p = .02$, $\eta^2 = .09$: police apps shared significantly more information (or provided document services) than third-party apps, although they only shared an average of 2.19 items out of a possible six.

[Insert Table 1. Host and App Features here]

### Surveillance or Tracking

A fisher's exact test revealed no relationship between *host* and *surveillance or tracking* systems, $\chi^2 = 0.25$, $p = .62$: police apps were no more or less likely to include surveillance mechanisms (5 apps, 10.42%) than third-party apps (1 app, 6.25%), although they could incite vigilantism.

### Navigation or Presentation

We also tested for relationships between host, and navigation or presentation and found that apps hosted by police or third-parties were no more or less likely to be well presented or easy to navigate. Overall, apps scored just above average for navigation (range = 0-2, $M = 1.09$, $SD = 0.83$), and slightly better for presentation (range = 0-2, $M = 1.31$, $SD = 0.75$), indicating that app design and maintenance could be improved.

### Anonymity, and Inclusivity, Reassurance, or Navigation

When testing the measures above, we noticed that the presence or absence of *anonymous reporting* options was related to three measures: *reassurance*, $F(1,62) = 6.22$, $p = .02$, $\eta^2 = .09$, apps that included anonymous reporting options also tended to include more reassuring items than apps with no anonymous reporting options; *inclusivity*, $\chi^2 = 12.64$, $p < .001$, apps with anonymous reporting mechanisms were more likely to be fully-inclusive than apps with no anonymous reporting options; and *navigation*, $\chi^2 = 7.78$ $p = .02$, apps with anonymous reporting mechanisms tended to be easier to navigate than apps with no anonymous reporting options.

[Insert Table 2. Anonymity, and Reassurance; Inclusivity; Navigation here]

Thus, hosts that included anonymous reporting options also appeared to be taking user experience seriously.

### Predicting User Ratings

Having evaluated the extent to which the apps *sought* to meet users' needs, while adhering to the vision of the hosts. We attempted to evaluate whether these measures were *actually* related to user experience. It is beyond the scope of this review to assess user experience in depth, which warrants exploration in future research, so, we simply tested whether the measures predicted user ratings. The null hypothesis was that the presence or absence of these measures was unrelated to user ratings.

Multiple linear regression was conducted using backward data entry. The variables of interest were *inclusivity*; *informed consent*; *anonymity; reassuring; safeguarding; empowering; engaging; information sharing; navigation; presentation; and host.* Seven of the models predicted user ratings. The best model included *empowering; information sharing; navigation* and *presentation*, $F(4,50) = 4.80$, $p < .001$ and explained 29.5% of variance in user ratings. Apps that were easy to navigate (e.g. without confusing links), that looked presentable (e.g. balance of text and images), were empowering (e.g. allowed users to advocate), and that shared information (e.g. statistics), were rated higher than apps that were not easy to navigate or presentable, and that did not empower or share information.

## Discussion

The current study aimed to evaluate and describe existing police/citizen communication apps across measures related to the mission of the Policing Vision 2025 (NPCC 2015) of making communities safer; and the US police principles of community policing (U.S. Department of Justice n.d.), particularly in relation to the role of police also as citizens and citizens as police (The Home office n.d., The Law Enforcement Partnership n.d). This was done by scoring apps (hosted by police or third parties), on the extent to which they included data protection and app features. It also briefly investigated how users rated these measures.

Just under 27% of apps were accessible without registering or providing login details. Some hosts collect data on users to monetise it, by designing the app to have user accounts, and this was more common in third-party apps. Also, despite the fact that 86% included an integrated reporting mechanism, only about half explicitly claimed to allow for anonymous reporting *prior to* reporting. More promising was the finding that inclusivity was related to anonymity, as fully inclusive apps were more likely claim to have anonymous reporting options than partially inclusive apps (as mentioned previously, we could not determine whether an app that claimed to provide anonymous reporting options actually did so in practice). Thus, a greater number of hosts of fully-inclusive apps also appeared to be aware that data collection of this sort could compromise anonymity when reporting. Hosts that claimed to have provided anonymous reporting options were also more likely to seek to reassure users, and to make their apps easier to navigate. Therefore, these hosts appeared to align attempts to *reassure* users that they were trustworthy with *actions*, by claiming to give users anonymous reporting options.

However, several apps offered 'anonymous' reporting while also collecting data that could potentially identify users. It seemed that users may not be aware that their anonymity could be compromised by providing e.g. location data[7]. Given the sensitive nature of reporting crimes or tips, anonymity was of concern when using several apps. Most notably, although police apps were generally more inclusive than those hosted by third-parties, 26 US police apps allowed for anonymous reporting, while 14 did not. Also, some tracked IP addresses covertly, some overtly, and others not at all, demonstrating the lack of consistency between police apps within just one country (discussed further below). Thus, while investigating anonymous reporting options in apps was revealing, anonymity cannot be reduced to this one domain, as other forms of data collection can compromise it. The main point is that users should know before using an app what will happen to their data and how much control they will have over it (see Perera et al. 2016, for guidelines for protecting privacy).

In the UK it is currently rare for a victim to report a crime against them anonymously and have it recorded as such (see National Crime Recording Standards, section 3.6., n.d.). However, anonymous digital reporting is useful for encouraging descriptive reporting of workplace harassment (e.g. Talk To Spot n.d., Vault n.d.) and mental health (e.g. Woebot), where people are often reluctant to speak to a human interviewer. Therefore, while anonymous victim crime reporting is currently rarely supported in the UK, it might be worth exploring as both a way of getting people to report crimes in the first place, and to provide sensitive details that they might not wish to disclose to a police officer (e.g. in cases of sexual assault). This should be balanced with the need to address risk and safeguard victims. Thus, anonymous reporting mechanisms should include two-way dyadic follow ups, where risks can be assessed and victims can be encouraged to provide details that they might have withheld in the initial report. The main point here is that users should have control over their anonymity in the initial stages of reporting, which can be modified later. Dyadic follow-ups could also be effective in filtering out the minority of false reports (estimated to be about 4% of cases by Kelly et al. 2005. See also Saunders 2012, for a review of rape allegations).

---

[7] Several apps tracked the IP address, email address, or location of the authors, sometimes without them knowing until they tried to access a function and were blocked, although no informed consent had been provided for the app to access these data. As data tracking issues emerged randomly and unexpectedly during analysis, the number of apps with these issues was not calculated.

Privacy policies are one way of acquiring informed consent. Therefore, for the present research, informed consent was measured by inspecting them – to see if they were accessible and easy to understand. For this, the privacy policy found on the landing page of each app was assessed for readability. About a third of apps either provided no privacy policy or there was no access to it (e.g. a broken URL), and Flesch reading ease labels revealed that only 10% were understandable to users without at least a university level education. Therefore, even when apps warned users about the ways they collected and used personal data, these warnings were generally confusing. More concerning was that privacy policies provided by police were no easier to read than ones provided by third-parties. If anything, there was a trend in the other direction. Only one of the three most readable privacy policies was provided with a police app (We Are All Police); the other two were provided with third-party apps (Rutland Neighbourhood Watch, and Your999). The apps analysed in the present research provided similar privacy agreement information and options as other apps more generally (e.g. yes/no buttons for access to location data). However, given civilian mistrust in police (Ray et al. 2017), it is worth making policing app functions transparent to users (see Brucato 2015, Jackson 2015), and giving users clear information about their anonymity and privacy before installing the app, and before using functions that need specific permissions.

One approach might be to consider privacy 'labels'. Kelley et al. (2009) found that informed consent was obtained more quickly and accurately, and the process was more enjoyable, when using privacy labels rather than policies. Therefore, for police/citizen communication apps, privacy labels might help to protect users better than privacy policies, as they should make it easier for users to understand how their data will be used.

Seeking trust was evaluated further by the measure 'reassuring users'. Given the association between police fairness and/or perceived legitimacy, and citizens' willingness to cooperate with police (Jackson et al. 2012), it was anticipated that apps hosted by police would be vigilant about including reassuring items, in an attempt to foster trust in a digital domain. However, apps hosted by police were no more likely to include these than apps hosted by third-parties. As discussed above, the only significant relationship found was related to anonymity, as apps that sought to reassure users with words (e.g. mission statements), also acted in a trustworthy way by seeking to protect them (anonymous reporting).

The second measure was top down safeguarding (including e.g. safety tips). Police apps included more safeguarding items than third-party ones. Thus, apps hosted by police generally aligned with the mission of the UK Policing Vision 2025 (NPCC 2015) of making communities safer, even though most of them were hosted by US police. In the US, the American Civil Liberties Union (n.d.) states that empowering citizens is important if they are to hold authorities to account, but a challenge facing policing is how to balance safeguarding and empowerment. Nevertheless, some apps successfully included both safeguarding items and ways to empower citizens (e.g. neighbourhood watch). Indeed, third-party apps included significantly more empowering items than police apps. However, some included items of concern. While several US police apps included useful sex offender information (e.g. Megan's Law 1996), some included live sex offender crime maps. This may have been an attempt at empowering citizens, but could also be seen to incite vigilantism. Indeed, Cubellis, et al. (2018) created a Sex Offender-Vigilante database that included 279 incidents of vigilantism against sex offenders (including murder), which shows that vigilantism is a real problem when citizens have tools to track individuals. Thirty-seven US police apps did not include sex offender maps and five did, demonstrating the lack of consistency in approaches to digital policing in the US, which might be related in part to the devolved governance of policing in the US (U.S. Department of Justice n.d.).

Digital technologies can also involve citizens in police-related activities or content e.g. 'web sleuthing' (Yardley et al. 2018). Many of the apps attempted to benefit from this by seeking to engage citizens, and police apps included the greatest number of engaging items (e.g. community events). The items appeared to be ways of fostering face-to-face contact between police and the community – which is important for reducing prejudice and increasing trust (Pettigrew 1998). Thus, when it came to engagement, police appeared to design apps to foster physical community contact. This suggested that they were trying to present themselves part of the community. In line with Bradford's work (2014), there may have been motivated to create a shared identity between police and users (citizens) – to encourage cooperation, disclosure, or social solidarity (Reicher and Hopkins 2000). However, as the items largely focused on presenting ways that police work *for* the community (e.g. 'rides-along'), it is unlikely that the items would have achieved a shared identity of police as citizens and citizens as police (The Home Office n.d., The Law Enforcement Partnership n.d.). This was also hinted at by the fact that police and third-party apps prioritised empowerment and engagement differently. Police apps were less focused than third parties on empowering the public (potentially out of a concern about control) and more focused on police-led engagement. In short, police apps appeared to be designed more in line with the UK Policing Vision 2025 (NPCC 2015) of protecting and safeguarding citizens, than the Peelian principle of seeing police as citizens and citizens as police (The Home Office n.d., The Law Enforcement Partnership, n.d) and encouraging the police and citizens to see this also.

Police have traditionally shared one-way information with citizens (Heverin and Zach 2010), but technology offers the possibility of two-way interactions. So, we explored two-way information sharing interactions. We were under the impression when starting this research that two-way interactions were common in citizen-police apps, as many landing pages advertised integrated feedback and reporting mechanisms, but this was not the case in reality. Police apps contained significantly more information sharing items than third-party apps, but information was almost exclusively top-down, and primarily about disseminating crime, incident, traffic information, or news (see Heverin and Zach), although statistics, reports and FAQs were also found. Other than many apps offering a one-way reporting mechanism and a few including a one-way user feedback function, no police app allowed citizens to contribute to information sharing, or question or co-create content. Also, no police app allowed users to see feedback of other users, despite this being recommended by Six (2003) as a way of fostering trust. This may be by design (to control content), but it might be worth considering mechanisms that allow for bottom-up information sharing (e.g. incident updates), or dashboards containing citizen feedback and actions taken, to encourage two-way communication or to allow "digital citizens" to participate in and challenge the police practices (Mossberger et al. 2007). This has become particularly apparent since the killing of George Floyd, both in the US (BBC 2020) and in the UK (Koram 2020). It seems that apps are not capitalising on the potential of digital technology to allow for truly dyadic communication. This means that citizens have no way of accessing, for instance, responses to feedback of others, and have no way of directly contributing to content meaningfully.

The analyses described above investigated the extent to which app hosts had *sought to* include appropriate measures. So, as a final step, we tested whether the measures were related to user experience, using user ratings as a measure of experience. We found that *empowering; information sharing; navigation;* and *presentation* were most significantly associated with user ratings, indicating that users are most influenced by the app's appearance or usability, and the degree to which the app informed or empowered them, when rating it. This suggests that it is worth making well-constructed and well-maintained apps that include both top-down information and means to empower users, as this may provide a more holistic experience for digital citizens. However, user ratings are a limited way of making these determinations, which would be tested more effectively experimentally or with qualitative analyses of user comments. These approaches warrant further investigation in future research.

There were also other limitations to the study. First, while the work is relevant to the design of future police apps internationally, this paper is pitched at US and UK policing, as the authors of the present research are based in the UK, and were limited to apps that were available in the UK version of Google play. While the apps were hosted in at least 34 countries (the countries of nine apps were unknown) most were US apps (78 of the 171 that we installed), of which 55 were hosted by police. Only ten were UK or UK Overseas Territories apps, of which two were hosted by police. A police practitioner also informed us (anecdotally) that he had never known of a report originating from a UK based third-party app, and that apps are seldom used to report crimes in the UK, although several online tools have recently been launched for people to report Covid-19 lockdown breaches (Burgess 2020). Second, our decision not to use the reporting mechanisms was taken to avoid submitting fake reports, but this limited our understanding of the ways that they worked in reality. For instance, we were only able to measure anonymity according to explicit statements about reporting anonymity prior to reporting, rather than whether reports were anonymous in reality. We were also not able to determine whether reports could be made (and saved) without submitting them straightaway, or what feedback/updates looked like. Thus, this study looked at the extent to which apps *sought* to provide various measures, rather than whether this was achieved in reality.

To summarise, our analyses focused on comparing apps hosted by police with those hosted by third parties. It became apparent that police apps appeared to take a top-down rather than collaborative role. They tended to focus on police-led engagement, top-down safeguarding and information-sharing, rather than on reassuring or empowering, while third-party apps generally focused on empowering. For instance, in police apps, there was seldom a means for users to co-create content, and information was filtered. We were also surprised to see a lack of steps taken to foster trust, and had concerns about how they gathered informed consent. First, because several apps tracked IP addresses, sometimes covertly, and second, because even when apps included a privacy policy before installation, they were largely very difficult to understand. As there is mistrust in police, police should be finding ways to tackle it digitally, or they risk amplifying it. Creating digital accountability mechanisms with measures to safeguard privacy and anonymity might go some way towards achieving this.

However, there was also some inconsistencies. For instance, the two US police apps were the highest scoring, while two were the lowest scoring. Some differences were justified (e.g., level of information provided), whereas others were problematic (e.g., provision of anonymous reporting). This suggests either that there could be a disconnect between the concept of having a policing app, its architecture, and the police officers who manage and use it (e.g. Sanders and Henderson 2013); or that there is a lack of a consistent digital policing message (U.S. Department of Justice n.d). Another explanation might be related to the devolved governance of policing down to local level in the US (U.S. Department of Justice n.d.), compared with the UK, where solutions like the "Single Online Home" have been developed centrally. It is also worth bearing in mind that external app developers have their own interests, so app design will not always be driven solely by policing concerns. Nevertheless, more needs to be done to agree upon a coherent message *before* making digital mechanisms, as these can amplify inconsistencies or underlying issues (e.g. bias) and potentially undermine policing efforts. Taking all this into account, while community policing might not be the focus of existing apps, building on the work of Cordner and Perkins (2013), carefully crafted digital communication systems between citizens and police have the potential to contribute to community policing in a digital world.

Based on the systematic evaluation of these apps and our discussion points, we therefore propose twelve design considerations for those developing digital mechanisms (as stand-alone apps or integrated into websites) for community-police collaboration. Two and three are exclusively for mechanisms that include reporting, while the others are considerations for all digital community-police mechanisms.

1. Digital mechanisms should provide privacy policies that are readable (this can be checked by using the Flesch online calculator (n.d.)) and concise. These should be clearly provided on the landing page, and within the mechanism itself (particularly before reporting), so that users can trust that their data will only be used in ways to which they have provided *informed* consent.

2. Digital *reporting* mechanisms might consider giving anonymous reporting options that give the user control over what kind of personal data they consent to sharing when making an initial report (e.g. IP addresses), and control over who can see the report (e.g. specially-trained officers when dealing with sexual abuse reports). They should also consider two-way follow ups, so that risks to the user can be assessed, modifications to anonymity can be discussed, and so that false reports can be filtered out.

3. Digital *reporting* mechanisms should make it explicit prior to making a report exactly how to report and what will happen to a report, so that users know to what they are consenting when reporting. This can be achieved using contextual help features, demos or screenshots.

4. Digital mechanisms might consider including items that help users to understand the aims and objectives of police, and their rights. These could help to increase transparency and foster trust.

5. Digital mechanisms might consider including safeguarding items, so that users can protect and empower themselves, and members of the community.

6. Digital mechanisms might consider including items designed to empower users in positive ways, so that they can participate in and challenge authorities and institutions, they can keep themselves and communities safe, and they can become an active part of building transparency and trust.

7. Digital mechanisms should not include items that have the potential to incite vigilantism, such as sex offender mapping.

8. Digital mechanisms might consider including items designed to engage users, so that they can participate in, create, and contribute to content. This should help to create an identity with shared values, where police see themselves and are seen also as citizens, and citizens can contribute to policing. This could encourage intervention for the common good.

9. Digital mechanisms should share accurate and timely information with users, but take care not to filter it in a way that either patronises users or masks elements that are key to transparency and building trust. It might also be worth considering including elements that encourage two-way information sharing and feedback.

10. Digital mechanisms should be easy to navigate, with buttons that provide direct and clear links to functions, rather than navigation loops, broken URLs, or unnecessary links to web pages. We also recommend that they look simple, clear and uncluttered.

11. Police forces might consider working together to decide upon a coherent and consistent message when creating digital mechanisms, and include elements amplify this message rather than contradict it. However, the architecture of the digital mechanism should be flexible enough to be adapted to the concerns of the demographic that each mechanism serves.

12. Digital mechanisms should be maintained and monitored by trained teams. For those with reporting mechanisms, teams should include staff with training in handling victim and witness reports ethically.

In conclusion, we set out to investigate two-way digital communication between police and citizens. However, despite only selecting apps whose landing pages gave the impression of two-way communication, in reality, only a few allowed for truly dyadic communication. Thus, it appears that police and citizens do not yet communicate digitally in this way. When it came to comparing third party and police apps, third party apps tended to focus on empowering citizens, while police apps tended to position police as separate from citizens, focusing on safeguarding, sharing top-down information, and police-led engagement. However, surprisingly, police apps took no more steps to reassure users than third party apps, and their privacy policies were equally difficult to read, so they probably largely failed to attain informed consent from users. This is of particular concern when user data is being collected without users' knowledge, for instance if they think they are making an anonymous report, but their location is known to the app host. As trust in policing is key to citizen cooperation, and we are increasingly living digital lives, if police forces are considering providing or updating a digital mechanism to communicate with citizens, it would be worth ensuring that the mechanism is, and is seen to be, trustworthy.